\documentclass[showpacs,preprintnumbers,10pt,twocolumn]{revtex4}%
\usepackage{amssymb}
\usepackage{amsfonts}
\usepackage{amsmath}
\usepackage{graphicx}
\usepackage{dcolumn}
\usepackage{bm}
\usepackage{revsymb}%
\providecommand{\U}[1]{\protect\rule{.1in}{.1in}}
\begin{document}
\title{Comment on \textquotedblleft Directed motion of spheres by unbiased driving
forces in viscous fluids beyond the Stokes' law regime\textquotedblright}
\author{Pedro J. Mart\'{\i}nez$^{1}$ and Ricardo Chac\'{o}n$^{2}$}
\affiliation{$^{1}$Instituto de Nanociencia y Materiales de Arag\'{o}n (INMA),
CSIC-Universidad de Zaragoza, E-50009 Zaragoza, Spain and Departamento de
 F\'{\i}sica Aplicada, E.I.N.A., Universidad de Zaragoza, E-50018 Zaragoza, Spain}
\affiliation{$^{2}$Departamento de F\'{\i}sica Aplicada, E.I.I., Universidad de
Extremadura, Apartado Postal 382, E-06006 Badajoz, Spain, and Instituto de
Computaci\'{o}n Cient\'{\i}fica Avanzada (ICCAEx), Universidad de Extremadura,
E-06006 Badajoz, Spain}
\date{\today}

\begin{abstract}
In a recent paper by Casado-Pascual [Phys. Rev. E \textbf{97}, 032219 (2018)],
directed motion of a sphere immersed in a viscous fluid and subjected to
zero-average biharmonic forces is studied. The author explains the dependence
on the relative amplitude of the two harmonic components of the average
terminal velocity from the perspective of a general formalism. In this Comment,  this
explanation is shown to be  in general incorrect, while the theory of ratchet
universality together with the vibrational mechanics approach provide a
satisfactory explanation of major aspects of the observed phenomena.

\end{abstract}
\maketitle

Motivated by investigation into the effects of nonlinear dissipative (drag)
forces on directed ratchet motion in the absence of any periodic substrate
potential, Casado-Pascual [1] has studied theoretically and numerically
spheres immersed in a viscous fluid and subjected to time-periodic forces of
zero average. In dimensionless variables, the equation of motion reads%
\begin{align}
\omega\tau\frac{d\boldsymbol{\upsilon}\left(  \theta\right)  }{d\theta}  &
=-\frac{1}{24}C_{d}\left[  \left\vert \boldsymbol{\upsilon}\left(  \theta\right)
\right\vert \right]  \left\vert \boldsymbol{\upsilon}\left(  \theta\right)
\right\vert \boldsymbol{\upsilon}\left(  \theta\right)  +f_{0}\mathbf{f}\left(
\theta\right)  ,\tag{1}\\
\mathbf{f}\left(  \theta\right)   &  =\zeta\cos\left(  \theta\right)
\mathbf{e}_{1}+\alpha\left(  1-\zeta\right)  \cos\left(  2\theta
+\varphi\right)  \mathbf{e}_{2}, \tag{2}%
\end{align}
where $\theta\equiv\omega t$, $\tau\equiv m/\left(  6\pi r\eta\right)  $ is a
characteristic timescale, $\boldsymbol{\upsilon}\left(  \theta\right)
\equiv2r\rho_{f}\mathbf{v}\left(  \theta/\omega\right)  /\eta$, $C_{d}\left[
\left\vert \boldsymbol{\upsilon}\left(  \theta\right)  \right\vert \right]  $ is
the steady drag coefficient, $f_{0}$ is a dimensionless parameter accounting
for the strength of the driving force, $\mathbf{e}_{1}$ and $\mathbf{e}_{2}$
are two mutually perpendicular versors, $\zeta\in\left[  0,1\right]  $ and
$\varphi\in\left[  0,2\pi\right]  $ account for the relative amplitude and
initial phase difference of the two harmonic components, respectively, while
the value $\alpha=1$ is the only one considered in Ref.~[1]. In the adiabatic
limit $\left(  \omega\tau\ll1\right)  $, the average terminal velocity,
$\overline{\mathbf{V}}_{ad}$, is given by the integral%
\begin{equation}
\overline{\mathbf{V}}_{ad}=\frac{\delta_{0}^{2}\eta}{16\pi r\rho_{f}}\int
_{0}^{2\pi}d\theta\left[  \sqrt{1+\frac{4\sqrt{f_{0}\left\vert \mathbf{f}%
\left(  \theta\right)  \right\vert }}{\delta_{0}}}-1\right]^2  \frac
{\mathbf{f}\left(  \theta\right)  }{\left\vert \mathbf{f}\left(
\theta\right)  \right\vert }, \tag{3}%
\end{equation}
with $\delta_{0}=9.06$ [1].

Commenting on Fig.~3 of Ref.~[1] (second component of the dimensionless
average terminal velocity versus $\zeta$), the author claims that:
\textquotedblleft The curves in Fig.~3 also reveal that, for fixed values of
the other parameters, there exists an optimal value of $\zeta$ which maximizes
the second component of the average terminal velocity. Furthermore, as
$\omega\tau$ increases, the maximum velocity decreases and its location shifts
toward lower values of $\zeta$. It should be noted here that, in the lowest
order, the general formalism developed in Refs.~[27,28] [for Refs.~[2,3]]
leads to the approximate expression $\overline{V}_{2}\left(  \zeta\right)
\approx C\zeta^{2}\left(  1-\zeta\right)  $, where $C$ is independent of
$\zeta$. This expression vanishes at $\zeta=0$ and $\zeta=1$, and displays a
maximum at $\zeta=2/3$, thus qualitatively resembling the behavior seen in
Fig.~3. However, it is unable to account for the dependence of the location of
the maximum velocity on $\omega\tau$. This deficiency is not surprising, given
that the above approximation is expected to be accurate only for small values
of $f_{0}$ and, in Fig.~3, we have taken $f_{0}=100$.\textquotedblright\ 

This Comment will question some of the above statements. For any $\alpha>0$,
we shall argue that the maximum velocity is reached for $\zeta=2\alpha
/(1+2\alpha)$ as predicted by the theory of ratchet universality (RU) [4-6].
Indeed, it has been demonstrated for temporal and spatial biharmonic forces
that optimal enhancement of directed ratchet transport is achieved when
maximally effective (i.e., \textit{critical}) symmetry breaking occurs, which
implies the existence of a particular universal waveform [4-6]. Specifically,
the optimal value of the relative amplitude $\zeta$ comes from the condition
that the amplitude of the odd harmonic component must be twice that of the
even harmonic component in Eq.~(2), i.e.,%
\begin{equation}
\zeta_{opt}=\zeta_{opt}\left(  \alpha\right)  \equiv2\alpha/(1+2\alpha).
\tag{4}%
\end{equation}
Note that this finding is in sharp contrast with the prediction coming from
the aforementioned general formalism [2,3], namely, that the dependence of the
average terminal velocity should scale as
\begin{equation}
\overline{V}_{2}\left(  \zeta\right)  \approx C\zeta^{2}\alpha\left(
1-\zeta\right)  . \tag{5}%
\end{equation}
Equation (5) indicates that $\overline{V}_{2}\left(  \zeta\right)  $ presents
again a single maximum at $\zeta_{opt}=2/3$, \textit{irrespective} of the
particular value of the prefactor $\alpha$. In contrast, both the theoretical
estimate given by Eq.~(3) and numerical simulations of Eq.~(1) confirm the
RU prediction [Eq.~(4)] over a wide range of $\alpha$ values (cf. Figs.~1 top
and bottom, respectively). \begin{figure}[ptb]
\includegraphics[height=0.45\textwidth,width=0.20\textwidth, angle=-90]{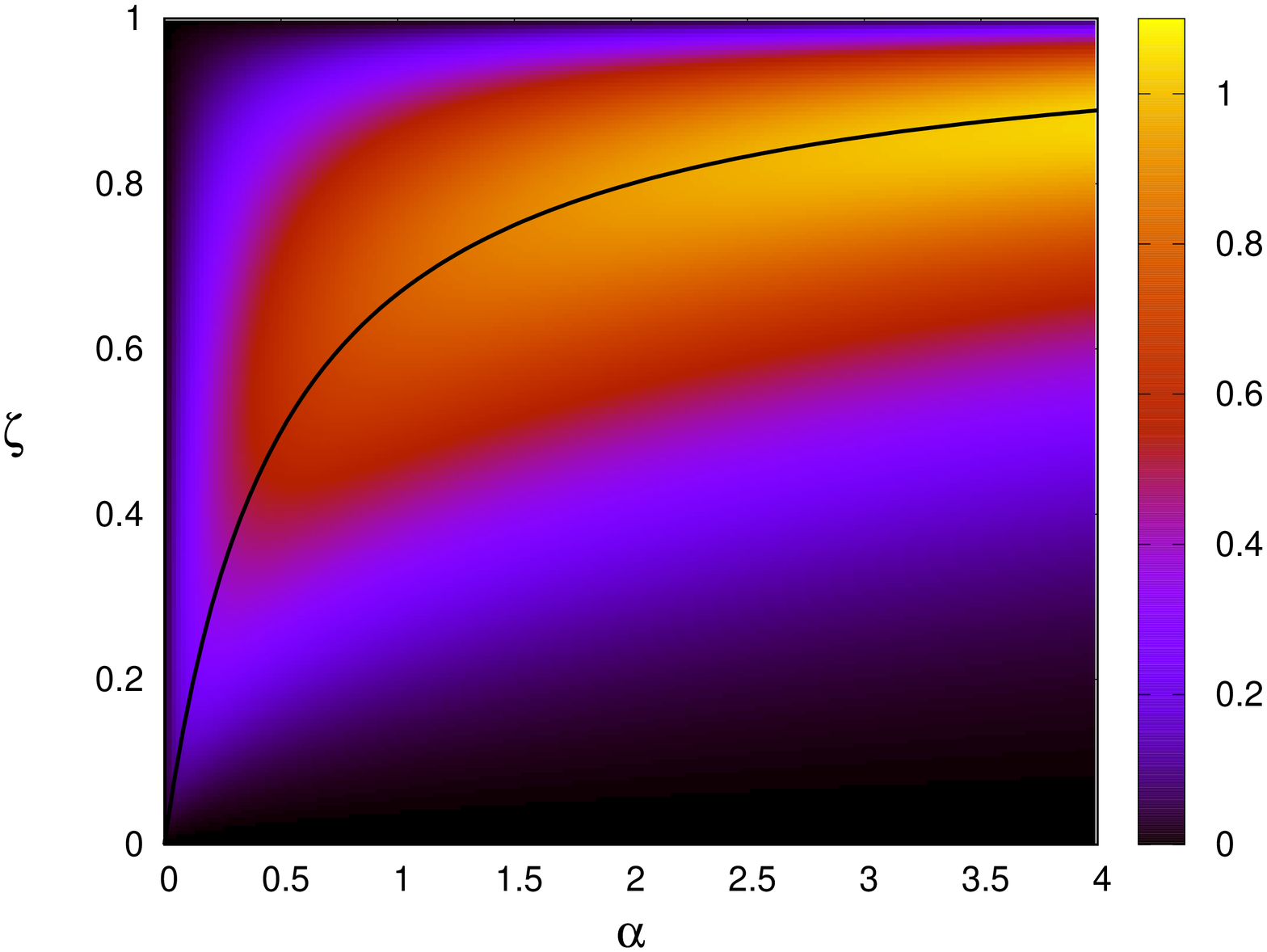}
\includegraphics[height=0.45\textwidth,width=0.20\textwidth, angle=-90]{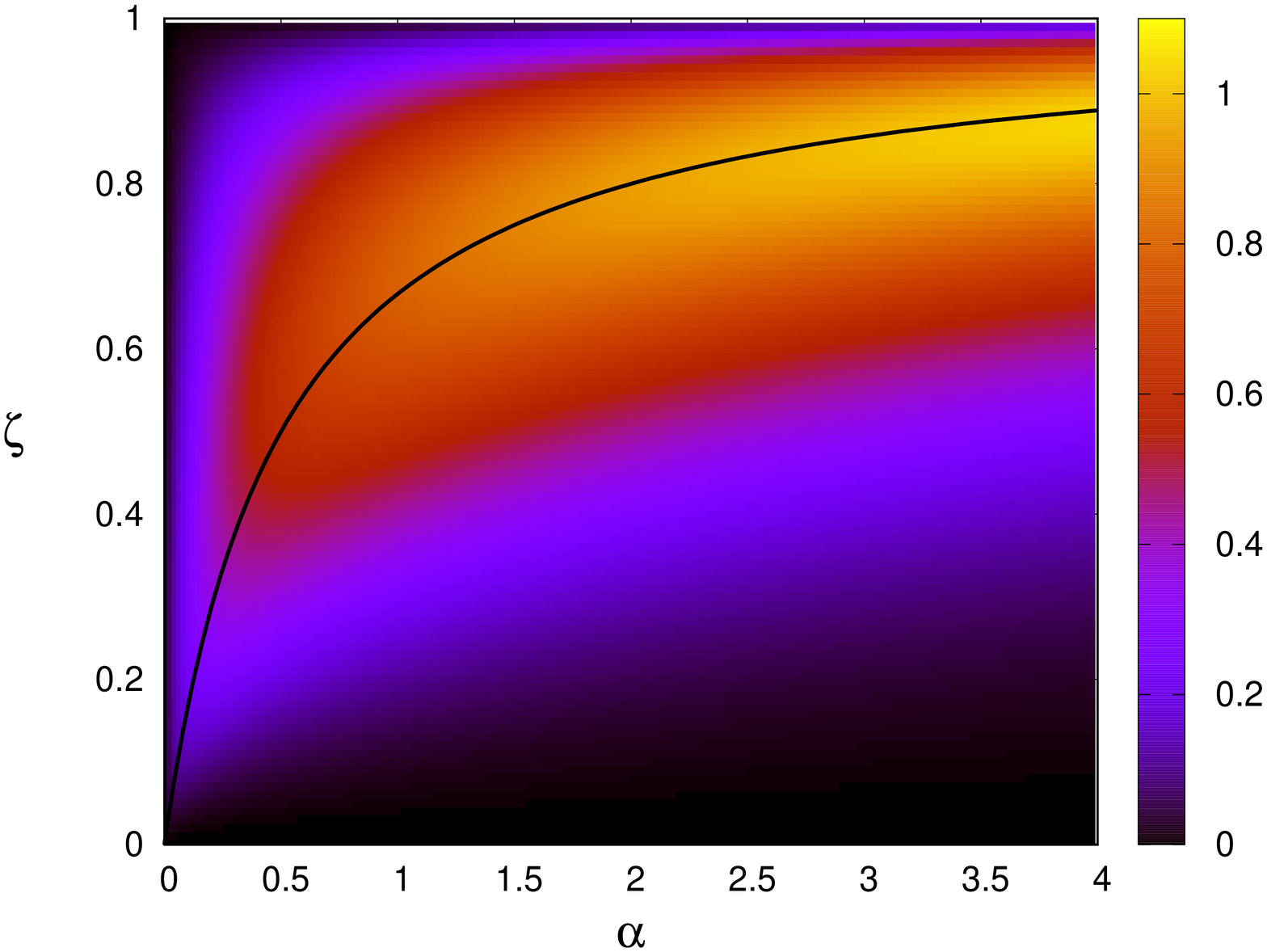}
\caption{Dimensionless average terminal velocity versus relative amplitude
$\zeta$ and prefactor $\alpha$ [cf. Eq.~(2)] for $f_{0}=100, \omega\tau=0.1$,
and $\varphi=\pi$. Top: Theoretical prediction from Eq.~(3). Bottom: Numerical
results from Eqs.~(1) and (2). Also plotted is the theoretical prediction for
the maximum velocity [cf. Eq.~(4); solid curve].}%
\label{fig1}%
\end{figure}\begin{figure}[ptb]
\includegraphics[height=0.45\textwidth,width=0.2\textwidth, angle=-90]{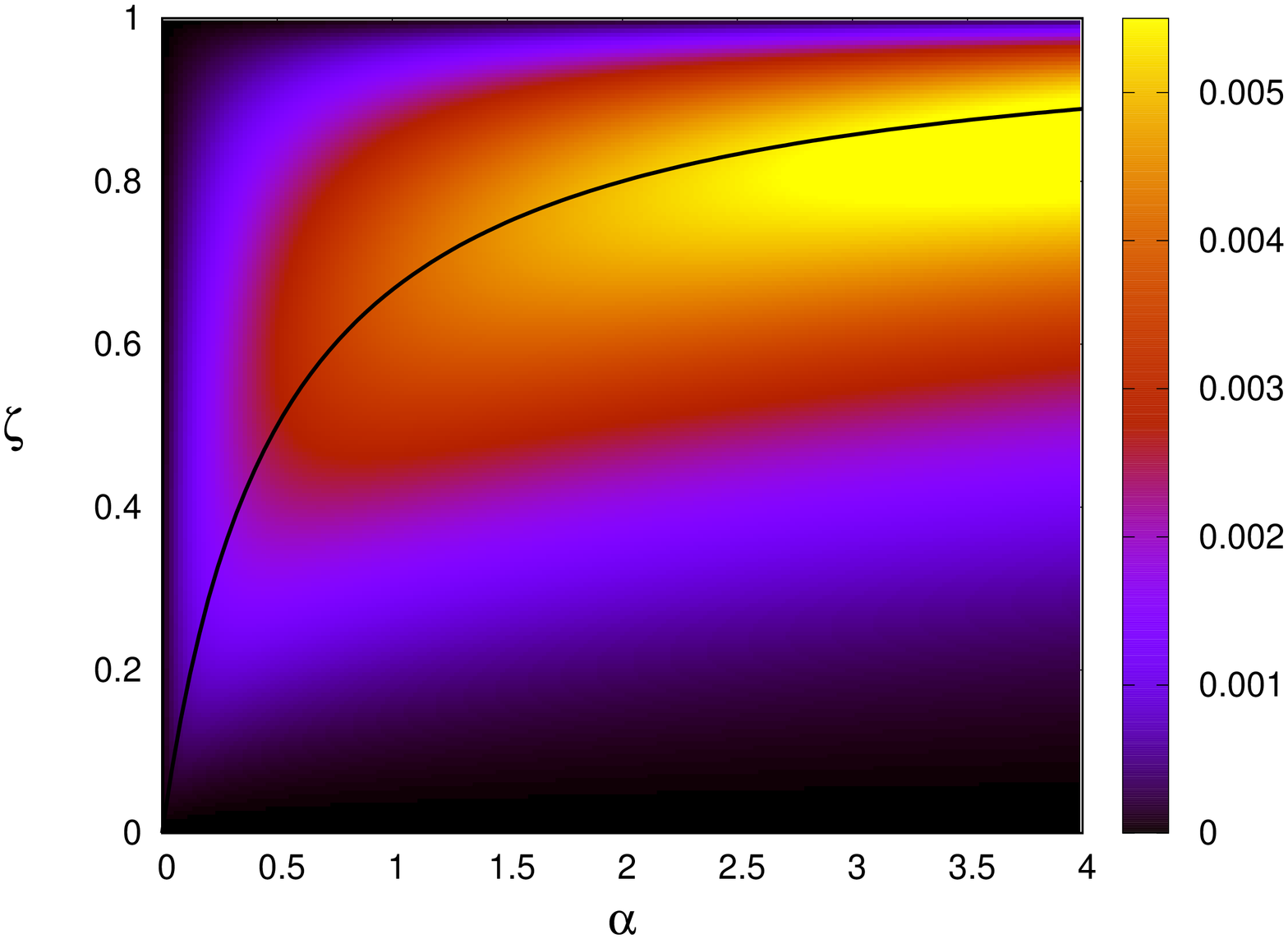}
\includegraphics[height=0.45\textwidth,width=0.2\textwidth, angle=-90]{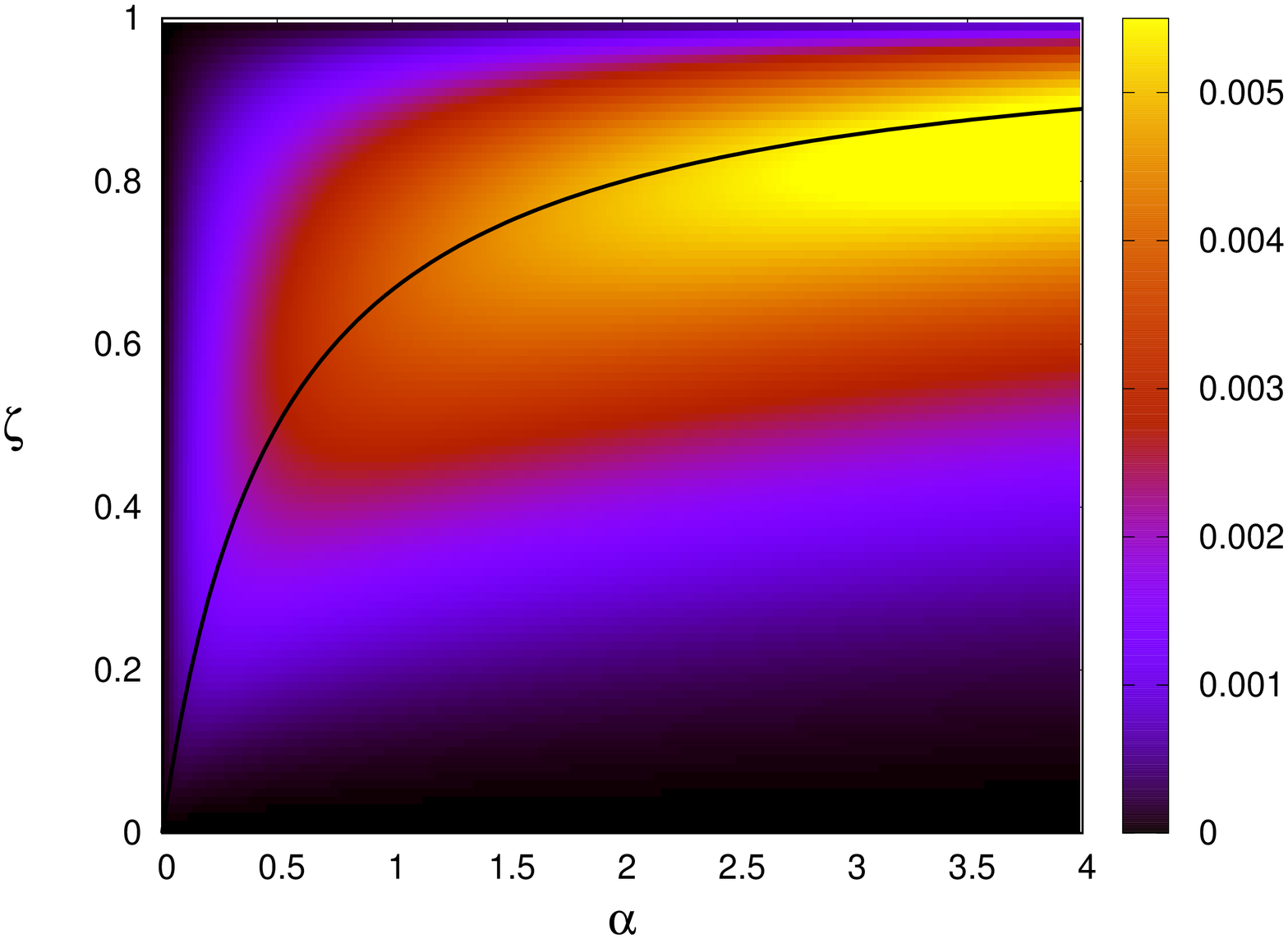}\caption{The
same as Fig.~1, but now $f_{0}=1$.}%
\label{fig2}%
\end{figure}

Figure 2 is the same as Fig.~1 but for a much lower dimensionless
driving strength $\left(  f_{0}=1\right)  $. One sees that the theoretical
estimate given by Eq.~(3) and numerical simulations of Eq.~(1) again confirm
the RU prediction [Eq.~(4)] over the same range of $\alpha$ values (cf. Figs.
2 top and bottom, respectively). Note that the average terminal velocity
increases as the prefactor $\alpha$ is increased, while keeping the remaining
parameters constant (cf. Figs.~1 and 2), because the condition $\left\vert
\mathbf{f}\left(  \theta\right)  \right\vert \leqslant1$ is no longer
satisfied for $\alpha>1$ and the particular value of $F_{0}$ considered in
Ref.~[1] for $\alpha=1$.

\begin{figure}[tbh]
\includegraphics[height=0.45\textwidth,width=0.2\textwidth, angle=-90]{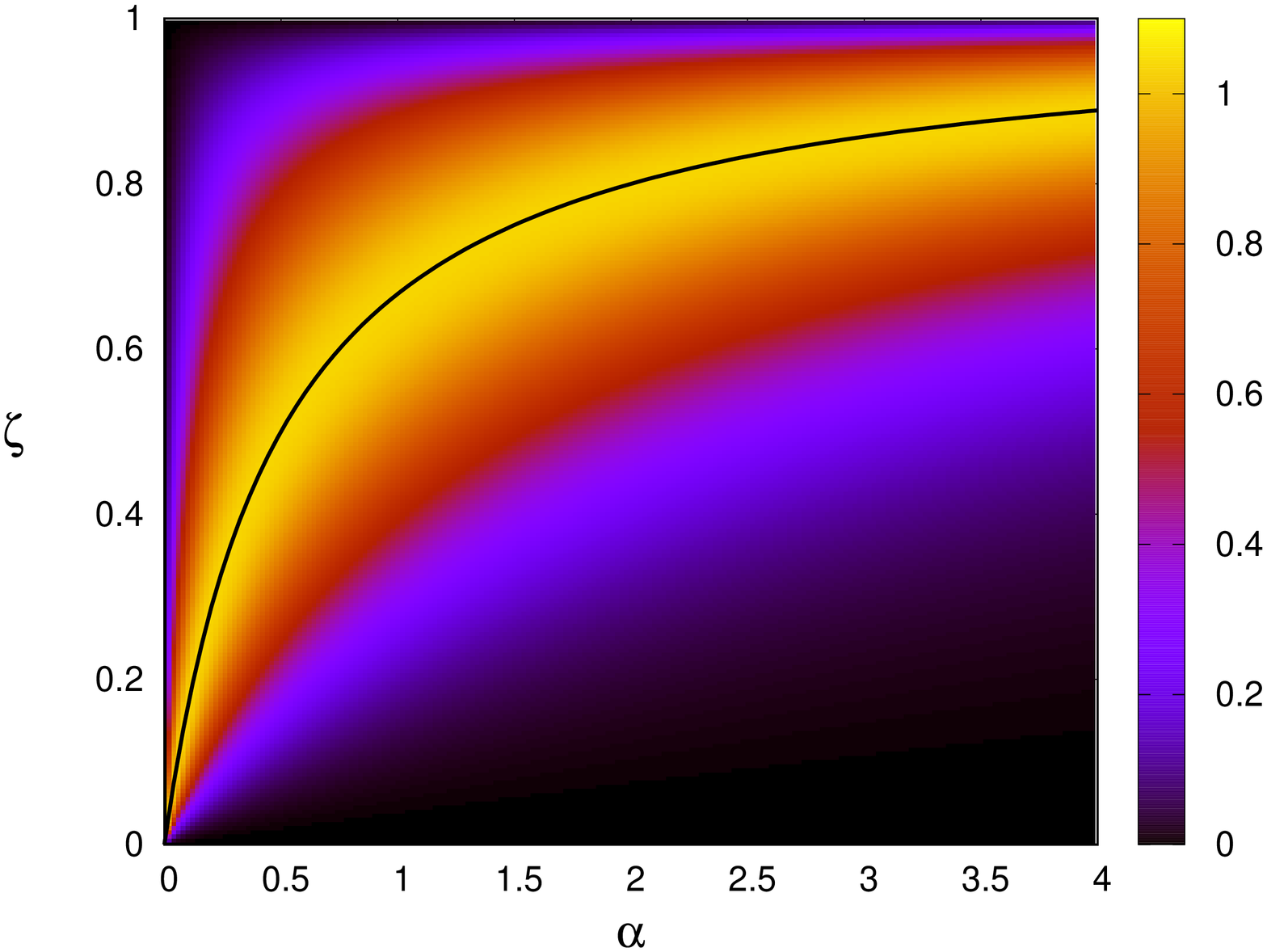}
\includegraphics[height=0.45\textwidth,width=0.2\textwidth, angle=-90]{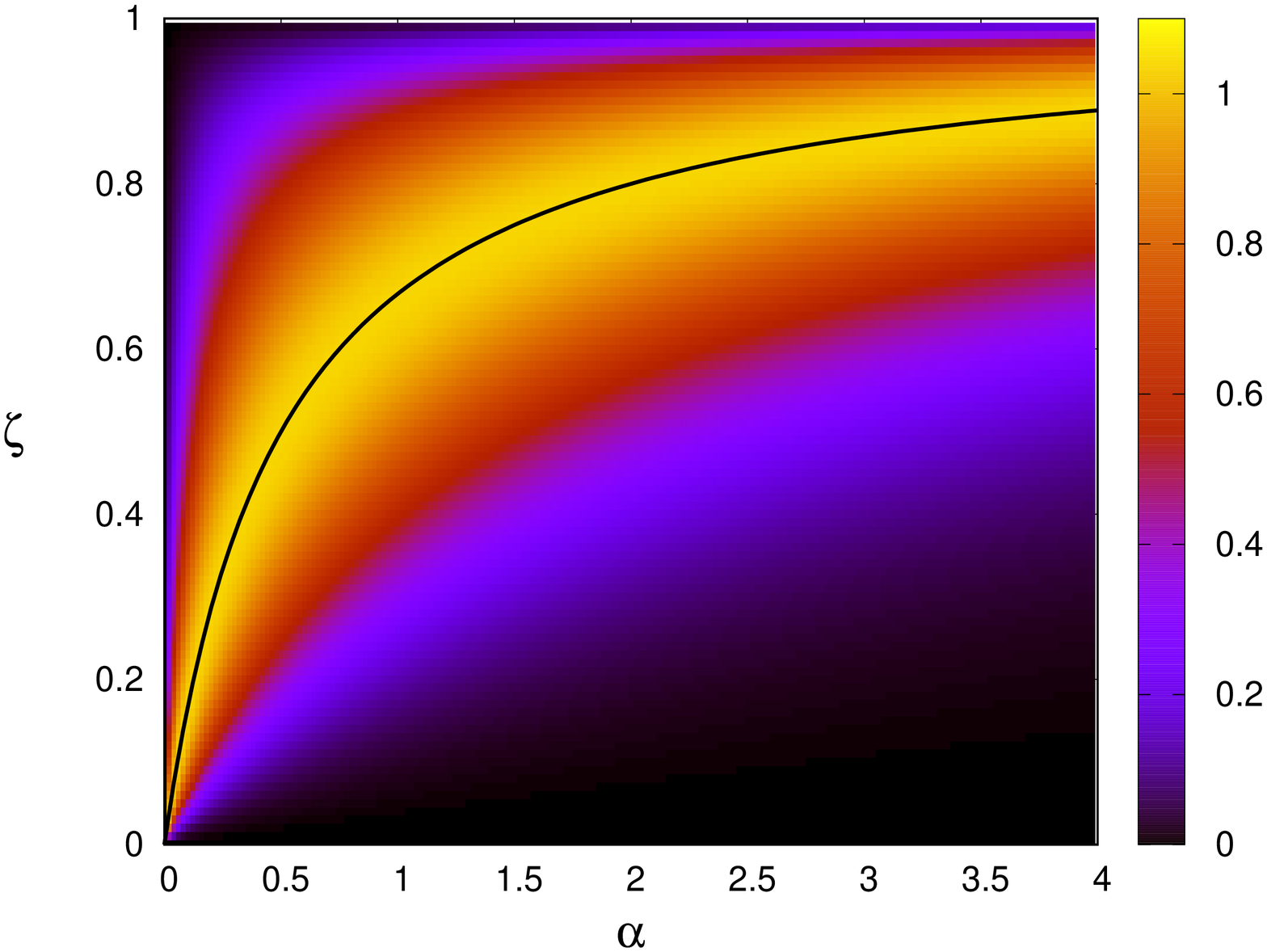}
\includegraphics[height=0.45\textwidth,width=0.2\textwidth, angle=-90]{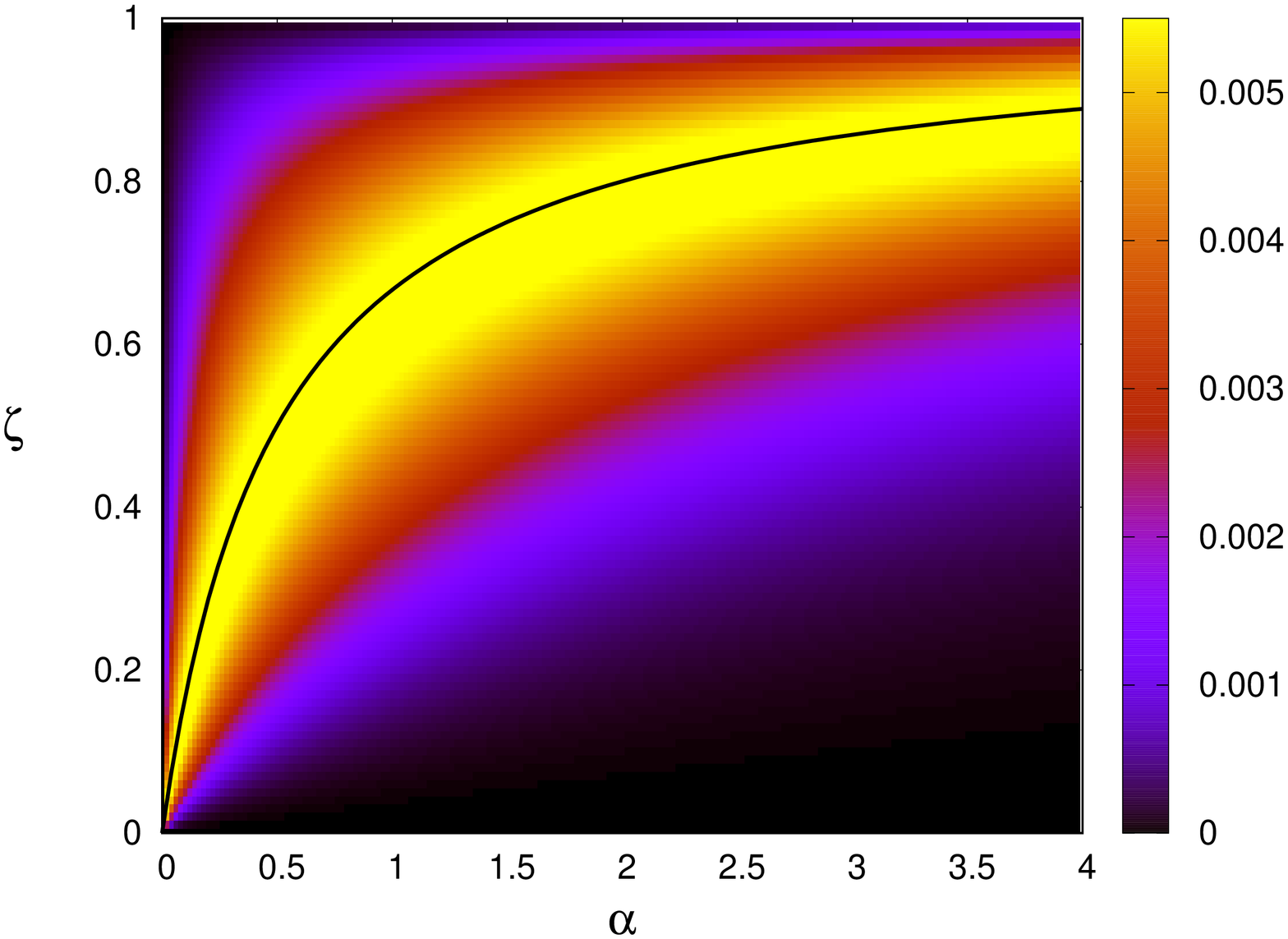}\caption{Dimensionless
average terminal velocity versus relative amplitude $\zeta$ and prefactor
$\alpha$ [cf. Eq.~(2)] for $\omega\tau=0.1$, $\varphi=\pi$, and $F_{0}$ such
that $\left\vert \mathbf{f}\left(  \theta\right)  \right\vert \leqslant1$.
Top: Theoretical prediction from Eq.~(3) for $f_{0}=100$. Middle: Numerical
results from Eqs.~(1) and (2) for $f_{0}=100$. Bottom: Numerical results from
Eqs.~(1) and (2) for $f_{0}=1$. Also plotted is the theoretical prediction for
the maximum velocity [cf. Eq.~(4); solid curve].}%
\label{fig3}%
\end{figure}

For the sake of making a legitimate comparison, Fig.~3 shows the corresponding
results after choosing the suitable value of $F_{0}\equiv\sqrt{\zeta
^{2}+\alpha^{2}\left(  1-\zeta\right)  ^{2}}$ such that $\left\vert
\mathbf{f}\left(  \theta\right)  \right\vert \leqslant1$. Clearly, the RU
prediction presents excellent agreement with the results from numerical
simulations for quite disparate values of $f_{0}$ (cf. Figs.~3 middle and 
bottom) as well as with the theoretical estimate [Eq.~(3); Fig.~3 top]. The
effectiveness of the RU prediction [Eq.~(4)] can be understood as follows. In
the adiabatic limit $\left(  \omega\tau\ll1\right)  $, after substituting Eq.
(11) of Ref.~[1] into the expression $C_{d}\left[  \left\vert \boldsymbol{\upsilon
}_{ad}\left(  \theta\right)  \right\vert \right]  \left\vert \boldsymbol{\upsilon
}_{ad}\left(  \theta\right)  \right\vert \boldsymbol{\upsilon}_{ad}\left(
\theta\right)  $ with the assumption $\boldsymbol{\upsilon}_{ad}\left(
\theta\right)  \approx\mathbf{f}\left(  \theta\right)  $, and Taylor- and
Fourier-expanding the nonlinear part of this friction force, for instance for
$\varphi=\varphi_{opt}\equiv\pi$, Eq.~(1) can be written as%
\begin{align}
\omega\tau\frac{dv_{1}}{d\theta} &  =-v_1-A\sum_{n=1}^{\infty}b_{2n-1}\cos\left[
\left(  2n-1\right)  \theta\right]  +f_{0}\zeta\cos\theta,\tag{6}\\
\omega\tau\frac{dv_{2}}{d\theta} &  =-v_2-A\sum_{n=0}^{\infty}a_{2n}\cos\left(
2n\theta\right)  -f_{0}\alpha(1-\zeta)\cos\left(  2\theta\right)  ,\tag{7}%
\end{align}
where $A\equiv A\left(  \delta_{0},f_{0}\right)  $ and the Fourier
coefficients $a_{2n}\equiv a_{2n}\left(  \zeta,\alpha\right)  $,
$b_{2n-1}\equiv b_{2n-1}\left(  \zeta,\alpha\right)  $ can be
calculated explicitly by using MATHEMATICA, but their size and algebraic complexity
prevent us from showing them easily. One sees that the net periodic force in
Eq.~(6) only presents odd harmonics and hence satisfies the shift symmetry, 
which
means that, as expected from the symmetry analysis in Ref.~[1],
such a periodic force by itself cannot yield directed ratchet
motion in the
$\mathbf{e}_{1}$ direction.
Also, the net force in Eq.~(7) only presents even harmonics and a
constant force term%
\begin{equation}
-Aa_{0}\approx\sum_{k=0}^{\infty}\sum_{n=0}^{k}\frac{\binom{k}{n}\zeta
^{2k-2n}\alpha^{2n+1}\left(  1-\zeta\right)  ^{2n+1}c\left(  k,n\right)
}{\Gamma\left(  \frac{1}{8}-k+1\right)  k!\left[  \zeta^{2}+\alpha^{2}\left(
1-\zeta\right)  ^{2}\right]  ^{k}},\tag{8}%
\end{equation}
with \begin{widetext}
%\begin{align}
\begin{equation}
\frac{c\left(
k,n\right)}{\Delta_{k,n}}\equiv\sqrt{\pi}_{2}\widetilde{F}_{1}
\left(\frac{1}{2}, \frac{k+n+3}{2};n+\frac{3}{2};-1\right)+\left(
-1\right)  _{\ 2}^{n} \widetilde{F}_{1}\left(  \frac{k-n+2}{2}, \frac{k+n+3}{2};
\frac{k+n+4}{2};-1 \right)\Gamma\left(  \frac{k-n+2}{2}\right)  , \tag{9}
\end{equation}
\end{widetext}
where $\Delta_{k,n}\equiv\left(  -1\right)  ^{-n}\left[
\left(  -1\right)  ^{n}-\left(  -1\right)  ^{k}\right]  \Gamma\left(
n+1\right)  $ while $\Gamma\left(  \cdot\right)  $ and $_{2}\widetilde{F}%
_{1}\left(  \cdot,\cdot;\cdot;\cdot\right)  $ are the gamma and the
regularized hypergeometric functions, respectively. Since the waveform of any
pair of even harmonics $a_{2n}\cos\left(  2n\theta\right)  +a_{2n+2}%
\cos\left[  \left(  2n+2\right)  \theta\right]  $ in Eq.~(7) does not fit, for
any value of $\zeta,\alpha$, that of one of the four equivalent expressions of
the biharmonic universal excitation [4]%
\begin{align}
&  f_{0}\left[  \sin\theta\pm\frac{1}{2}\sin\left(  2\theta\right)  \right]
,\tag{10}\\
&  f_{0}\left[  \cos\theta\pm\frac{1}{2}\sin\left(  2\theta\right)  \right]
,\nonumber
\end{align}
this means that the emergence of directed ratchet motion in the $\mathbf{e}%
_{2}$ direction is exclusively due to the constant force $-Aa_{0}$ which is
given approximately by Eq.~(8) for the parameters used in Fig.~3 of Ref.~[1].
Indeed, the RU prediction [Eq.~(4)] presents good agreement with the
theoretical estimate of the constant force arising from the Fourier analysis
of the net force [cf. Eq.~(8); see Fig.~4] because the universal biharmonic
waveform is effectively present to produce the term of constant force
\textit{once} $\mathbf{f}\left(  \theta\right)  $ has been suitably normalized
as in the criticality scenario leading to ratchet universality [5].
\begin{figure}[ptb]
\includegraphics[width=0.45\textwidth]{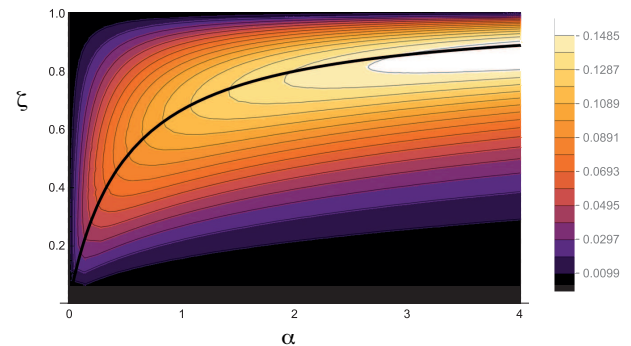}
\caption{Estimation of the constant term $-Aa_{0}$ after retaining only the
first six terms versus relative amplitude $\zeta$ and prefactor $\alpha$ [cf.
Eq.~(8)] for $f_{0}=100$. Also plotted is the theoretical prediction for the
maximum velocity [cf. Eq.~(4); solid curve].}%
\label{fig4}%
\end{figure}

Next, the aforementioned ``dependence of the location of the maximum velocity
on $\omega\tau$" (cf. Ref.~[1]) can be understood as follows. First, we
rewrite Eq.~(1) as%
\begin{align}
\frac{d\boldsymbol{\upsilon}\left(  \Omega t_{\tau}\right)  }{dt_{\tau}} &
=-\frac{1}{24}C_{d}\left[  \left\vert \boldsymbol{\upsilon}\left(  \Omega t_{\tau
}\right)  \right\vert \right]  \left\vert \boldsymbol{\upsilon}\left(  \Omega
t_{\tau}\right)  \right\vert \boldsymbol{\upsilon}\left(  \Omega t_{\tau}\right)
\nonumber\\
&  +f_{0}\left[  \zeta\cos\left(  \Omega t_{\tau}\right)  \mathbf{e}%
_{1}+\alpha\left(  1-\zeta\right)  \cos\left(  2\Omega t_{\tau}+\varphi
\right)  \mathbf{e}_{2}\right]  ,\tag{11}%
\end{align}
where $\Omega\equiv\omega\tau,t_{\tau}\equiv t/\tau$. For sufficiently large
$\Omega$, i.e., when $\tau\gtrsim T$, the $v_{2}$ dynamics of Eq.~(11) can be
analysed using the vibrational mechanics approach [7] by assuming that the
$\Omega$-force is ``slow" while the $2\Omega$-force is ``fast". Thus, one
separates $v_{2}\left(  t_{\tau}\right)  =V_{2}\left(  t_{\tau}\right)
+\psi\left(  t_{\tau}\right)  $, where $V_{2}\left(  t_{\tau}\right)  $
represents the slow dynamics while $\psi\left(  t_{\tau}\right)  $ is the fast
oscillating term: $\psi\left(  t_{\tau}\right)  =\psi_{0}\cos\left(  2\Omega
t_{\tau}+\Phi\right)  $ with $\psi_{0}\equiv f_{0}\alpha\left(  1-\zeta
\right)  /\sqrt{1+4\Omega^{2}}$, $\Phi\equiv\arctan\left[  \frac
{tg\varphi-2\pi}{1+2\Omega tg\varphi}\right]  $. On averaging out $\psi\left(
t_{\tau}\right)  $ over time $2\Omega t_{\tau}$, the slow reduced dynamics of
$v_{2}$ becomes%
\begin{equation}
\frac{dV_{2}}{dt_{\tau}}=-V_{2}+\frac{4\pi}{\delta_{0}}V_{2}^{3/2}.\tag{12}%
\end{equation}
Thus, the asymptotic $v_{2}$ dynamics when $\Omega\equiv\omega\tau\gg1$ could
well be described by Eq.~(12), which indicates that the (terminal) velocity
$V_{2}\sim e^{-t/\tau}$ as $t\rightarrow\infty$. This scenario is coherent
with the gradual decrease of the maximum second component of the dimensionless
average terminal velocity (cf. Fig.~3 in Ref.~[1]) as $\omega\tau$ is
increased from the adiabatic limit, i.e., as the relevant symmetries are
gradually restored. This phenomenon of competing timescales leads to the
$2\Omega$-force losing ratchet effectiveness, but without deactivating the
degree-of-symmetry-breaking mechanism [4,5]. Clearly, this loss of
effectiveness can only be compensated by increasing its amplitude
$\alpha\left(  1-\zeta\right)  $ to generate maximum velocity, thereby explaining
the shift of the location of the maximum velocity to lower values of $\zeta$
as $\omega\tau$ is increased from the adiabatic limit (cf. Fig.~3 in Ref.~[1]).

Finally, the author claims that: ``In this class of models [for rocking
ratchets in the presence of thermal noise], the mechanism behind the
generation of directed motion is basically harmonic mixing... ." This is
incorrect. It has been demonstrated that optimal enhancement of directed
motion is achieved when maximal effective (i.e., critical) symmetry breaking
occurs [4,5], while the effect of finite temperature on the purely
deterministic criticality scenario can be understood as an effective
noise-induced change of the potential barrier which is in turn controlled by
the degree-of-symmetry-breaking mechanism [8-10].

In conclusion, the theoretical explanation discussed in Ref.~[1] is in
general incorrect, while the theory of RU and the vibrational mechanics
approach provide a satisfactory explanation of major aspects of the observed phenomena.

P.J.M. acknowledges financial support from the Ministerio de Econom\'{\i}a y
Competitividad (MINECO, Spain) through Project No. FIS2017-87519 cofinanced by
FEDER funds and from the Gobierno de Arag\'{o}n (DGA, Spain) through Grant No.
E36\_17R to the FENOL group. R.C. acknowledges financial support from the
Junta de Extremadura (JEx, Spain) through Project No. GR18081 cofinanced by
FEDER funds.

\end{document}